\documentclass[12pt]{book}

\usepackage[dvips]{graphicx,color}
\usepackage{makeidx,tsukuba}
\makeauthorindex
\makeindex

\begin{document}

\BookTitle{\itshape The 28th International Cosmic Ray Conference}
\CopyRight{\copyright 2003 by Universal Academy Press, Inc.}
\pagenumbering{arabic}

\chapter{Observations of H1426+428 from 1999 to 2002 with The Whipple 
Observatory 10 m Telescope}

\author{%
D.~Horan,$^{1,2}$ I.H.~Bond, P.J.~Boyle, S.M.~Bradbury, J.H.~Buckley,
D.~Carter-Lewis, O.~Celik, W.~Cui, M.~Daniel, M.~D'Vali,
I.de~la~Calle~Perez, C.~Duke, A.~Falcone, D.J.~Fegan, S.J.~Fegan,
J.P.~Finley, L.F.~Fortson, J.~Gaidos, S.~Gammell, K.~Gibbs,
G.H.~Gillanders, J.~Grube, J.~Hall, T.A.~Hall, D.~Hanna, A.M.~Hillas,
J.~Holder, A.~Jarvis, M.~Jordan, G.E.~Kenny, M.~Kertzman, D.~Kieda,
J.~Kildea, J.~Knapp, K.~Kosack, H.~Krawczynski, F.~Krennrich,
M.J.~Lang, S.~LeBohec, E.~Linton, J.~Lloyd-Evans, A.~Milovanovic,
P.~Moriarty, D.~Muller, T.~Nagai, S.~Nolan, R.A.~Ong, R.~Pallassini,
D.~Petry, B.~Power-Mooney, J.~Quinn, M.~Quinn, K.~Ragan, P.~Rebillot,
P.T.~Reynolds, H.J.~Rose, M.~Schroedter, G.~Sembroski, S.P.~Swordy,
A.~Syson, V.V.~Vassiliev, S.P.~Wakely, G.~Walker, T.C.~Weekes,
J.~Zweerink \\ {\it (1) Smithsonian Astrophysical Obs., P.O. Box 97,
Amado, AZ 85645, USA}\\ {\it (2) The VERITAS Collaboration--see
S.P.Wakely's paper} ``The VERITAS Prototype'' {\it from these
proceedings for affiliations} }

\section*{Abstract}

The BL Lacertae object H1426+428 is the most distant, confirmed source
of TeV gamma rays. At a redshift of 0.129, its detection at TeV
energies has important implications for estimating the density of the
extragalactic infra-red background radiation. H1426+428 was observed
extensively during the 2001/2002 observing season with the Whipple 10
m gamma-ray telescope. The results of these observations are presented
here and are combined with the results of previous observations made
between 1999 and 2001 at Whipple.

\section{Introduction}

First discovered in the 2-6 keV band by HEAO 1 [12], and classified as
a BL Lacertae type object in 1989 [11], H1426+428 (H1426) has recently
been classified as an ``extreme'' blazar because the peak of its
synchrotron emission occurs at energies greater than 100 keV. Such
blazars are prime candidates for TeV emission if they have a
sufficient level of soft seed photons [5]. Indeed, H1426 was among
four blazars singled out by Costamante et al. [4] as likely TeV
emitters. This prediction was borne out when H1426 was detected in the
VHE band by Whipple in 2000 [8]. This detection was subsequently
confirmed by other ground-based atmospheric \v{C}erenkov experiments
[1,7] firmly establishing H1426 as a source of TeV gamma rays. Since
then, many observations of H1426 have been carried out at TeV
energies. Information about its energy spectrum have been derived
[1,10], and the implications that its detection has on the density of
the extragalactic IR background have been discussed [6,10]. In this
paper, the Whipple observations of H1426 between 1999 and 2002 are
summarised.

\section{Observations}

H1426 had been observed extensively with the Whipple telescope since
March 1999 when the first evidence for a VHE signal was seen
[8]. Throughout this time, the sensitivity of the Whipple instrument
varied due to changes in camera configuration, mirror reflectivity,
triggering conditions and pointing accuracy. The main characteristics
of the instrument during this time period are summarised in
Table~\ref{camera} A combined total of 110.5 useful hours of H1426
data were taken in both the PAIRS and the TRACKING modes [3] at
Whipple during the 2001/2002 observing season; these are summarised
in Table 2.

\begin{table}[t]
 \caption{\label{camera}Camera Configurations from 1999 to 2002}
\begin{center}
\begin{tabular}{l|cccc}
\hline
Observing              & Number    & Spacing  & FOV$^a$  & E$_{peak}$$^b$ \\
Season                 & of Pixels & [deg]    & [deg]    & [GeV] \\         
\hline
Mar. 1999 - Jun. 1999  & 331       & 0.24     & 4.8      & 500 \\
Sep. 1999 - Jul. 2000  & 490       & 0.12$^c$ & 3.8$^d$  & 430 \\
Oct. 2000 - Jun. 2001  & 490       & 0.12$^c$ & 3.8$^d$  & 390 \\
Oct. 2001 - Jul. 2002  & 490       & 0.12$^c$ & 3.8$^d$  & 400 \\
\hline
\end{tabular}
\end{center}
$^a$ Field of View [FOV]. \\
$^b$ The peak response energy; this is the energy at which the
collection area folded with an E$^{-2.5}$ spectrum reaches a maximum. \\
$^c$ The spacing between the outer tubes is 0.24$^\circ$.\\
$^d$ The outer ring of tubes was not used in this analysis, hence the
FOV here is effectively 2.6$^\circ$.
\end{table}

\begin{table}[t]
 \caption{\label{observations}H1426 Observations During the 2001/2002 Observing Season}
\begin{center}
\begin{tabular}{l|cc}
\hline
Observation Mode:                              & PAIRS  & TRACKING$^a$ \\
\hline
No. ON source scans taken:                     & 91     & 326    \\
No. ON source scans included in this analysis: & 73     & 247    \\
Total useful exposure [hrs]:                   & 33.3   & 110.5  \\
Total significance [$\sigma$]:                 & 2.1    & 2.4    \\
\hline
\end{tabular}
\end{center}
$^a$ All data taken ON source are included here, i.e. ON source
data taken in the PAIRS mode and data taken in the TRACKING mode.\\
\end{table}

\section{Results and Discussion}

The results of the H1426 observations between 1999 and 2001 have been
described in detail by Horan et al. 2002 [9] and are summarised here
in Table~\ref{results} The gamma-ray flux from H1426 during the
2001/2002 observing season was found to be weaker than during the
previous years. This is consistent with reports from the HEGRA
collaboration [2], who also found H1426 to be in a lower emission
state during this time period. The lightcurve for H1426 from March
1999 to July 2002 is shown in Figure 1. The rates are plotted for each
month during which H1426 was observed at Whipple and are expressed in
terms of the Crab rate for that observing season. The combined rate
for each period is also shown and can be seen to be lowest for the
observations taken during 1999 and 2002.

The average flux from H1426 in the soft X-ray band as recorded by the
All Sky Monitor was also found to be lower than in other years during
the time that H1426 was observed at Whipple in 2002. This is
consistent with results from other TeV blazars which reveal the
average X-ray and TeV gamma-ray flux levels to be correlated.

Due to its large redshift ($z$=0.129), spectral measurements of H1426
are of particular importance in determining the density of the
extragalactic infra-red background radiation. A detailed spectral
analysis by Petry et al. [10] revealed H1426 to have a spectrum steeper
than that of any other TeV blazar. The large dataset accumulated
during 2002 is being incorporated into this spectral analysis and an
update will be presented at the conference.

\begin{table}[t]
 \caption{\label{results}Results of H1426 Analysis from 1999 to 2002}
\begin{center}
\begin{tabular}{l|cccc}
\hline
Period of              & Exposure  & Total    & Max. $\sigma$ & F$_{peak}$$^b$\\
Observations           & [hrs]     & $\sigma$ & Month$^a$     & [x 10$^{-11}$ cm$^{-2}$ s$^{-1}$]\\
\hline
Mar. 1999 - Jun. 1999  &  24.35    & 0.9      & 1.6           & $<$ 0.2 \\
Feb. 2000 - Jun. 2000  &  26.37    & 3.1      & 3.4           & 0.35    \\
Jan. 2001 - Jun. 2001  &  31.12    & 5.5      & 5.0           & 0.88    \\
Jan. 2002 - Jul. 2002  & 110.54    & 2.4      & 2.7           & 0.30    \\
\hline
\end{tabular}
\end{center}
$^a$ The maximum statistical significance of the signal recorded from
H1426 when the data are grouped by the month during which they were
recorded.\\
$^b$ The integral flux above E$_{peak}$ for that year as given in Table 1.
\end{table}

\begin{figure}[t]
  \begin{center}
    \includegraphics[height=20pc]{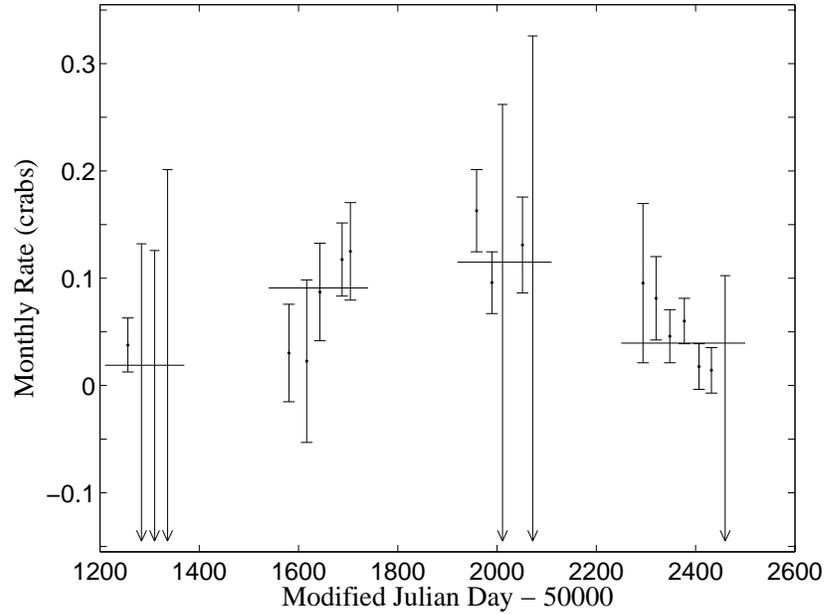}
  \end{center}
  \vspace{-0.5pc}

  \caption{\label{rate}The average gamma-ray rate from H1426 each
  month from 1999 to 2002. The combined rate for each of the four
  seasons is plotted as a horizontal line through each dataset. The
  rates are expressed in units of the Crab rate during that season.}

\end{figure}%

\section{Acknowledgments}

We acknowledge the technical assistance of E. Roache and J. Melnick.
This research is supported by grants from the U. S. Department of
Energy, by Enterprise Ireland and by PPARC in the UK. The ASM
quicklook results were provided by the ASM/RXTE team ({\it
http://xte.mit.edu}).

\section{References}

\vspace{\baselineskip}

\re
1.\ Aharonian F. A. \ et al. \ 2002, A\&A 384, L23
\re
2.\ Aharonian F. A. \ et al. \ 2003, A\&A {\it in press} astro-ph/0301437
\re
3.\ Catanese M. A. \ et al. \ 1998, ApJ 501, 616
\re
4.\ Costamante L. \ et al. \ 2000, Mem. Soc. Astron. Italia 72, 153
\re
5.\ Costamante L. \ et al. \ 2002, A\&A 384, 56 
\re
6.\ Costamante L. \ et al. \ 2003, A\&A  {\it in press} astro-ph/0301211
\re
7.\ Djannati-Atai A. \ et al. \ 2002, A\&A 391, L25
\re
8.\ Horan D. \ et al. \ 2001, AIP Conf. Proc. 587, 324
\re
9.\ Horan D. \ et al. \ 2002, ApJ 571, 753
\re
10.\ Petry D. \ et al. \ 2002, ApJ 580, 104
\re
11.\ Remillard R. A. \ et al. \ 1989, ApJ 345, 140
\re
12.\ Wood K. S. \ et al. \ 1984, ApJS 56, 507

\endofpaper
\end{document}